\documentclass[useAMS,usenatbib,onecolumn]{mn2e}
\usepackage{graphicx,amssymb,amsmath,times}
\title{Revisiting the effect of nearby supernova remnants on local cosmic rays}
\author[S. Thoudam ]{Satyendra Thoudam\thanks{E-mail: satyend@barc.gov.in}\\
Astrophysical Sciences Division,
Bhabha Atomic Research Centre, Trombay, Mumbai-400085,
Maharashtra, India}
\begin{document}
\date{}
\pagerange{}
\maketitle
\label{firstpage}
\begin{abstract}
In an earlier paper, the effect of the nearby known supernova remnants (SNRs) on the local cosmic-rays (CRs) was studied, considering different possible forms of the particle injection time. The present work is a continuation of the previous work, but assumes a more realistic model of CR propagation in the Galaxy. The previous work assumed an unbounded three-dimensional diffusion region, whereas the present one considers a flat cylindrical disc bounded in both the radial and vertical directions. The study has found that the effect of the vertical halo boundary $H$ on the local SNR contribution to the observed CR anisotropy is negligible as long as $H\gtrsim 2kpc$. Considering the values of the halo height $H\gtrsim 2kpc$ obtained by different authors, the present work suggests that the study of the effect of local sources on the CR anisotropy can be carried out without having much information on $H$ and hence, using the much simpler three-dimentional unbounded solution. Finally, the present work discusses about the possibility of explaining the observed anisotropy below the knee by a single dominant source with properly chosen source parameters, and claims that the source may be an \textit{undetected} old SNR with a characteristic age of $\sim 1.5\times 10^5 yr$ located at a distance of $\sim 0.57 kpc$ from the Sun.   

\end{abstract}

\section{Introduction}
There is a wealth of literature available on the highly isotropic nature of cosmic-rays (CRs) observed at the Earth (see e.g. the references given in Thoudam 2007, hereafter Paper I). The CR anisotropy amplitude is only $\sim (10^{-4}-10^{-3})$ in the energy range of $(10^{11}-10^{15})eV$ (Guillian et al. 2007 and references therein) with the phase (direction) mainly found in the outer Galaxy, particularly in the second quadrant of the Galaxy. The possible explanations for the anisotropy are generally beleived to be the global diffusion leakage of CRs from the Galaxy, the random nature of the CR sources in space-time and the effect of the local sources. In Paper I, the effect of the known local supernova remnants (SNRs) has been studied in detail by giving more emphasis to the particle release time. The study found that the observed anisotropy data favour the burst-like injection model if particles are released from the sources at an age of $\sim (2-5)\times 10^4 yr$. The continuous injection model gives an anisotropy which is too large to explain the observed data. However, Paper I considered the CR diffusion zone as an unbounded three-dimensional space which is actually too far from the real geometry of the Galaxy. The present work is a continuation of the earlier work, but considers the diffusion region as a flat cylindrical disc having both radial and the vertical boundaries. 

In the present study, the propagation of CRs is assumed to follow the same diffusion equation given in Paper I. The solution will be applied to local SNRs and the results  will be compared to those obtained in Paper I for the burst-like model of particle injection. 

\section {CR spectrum from a point source}

In the diffusion model, neglecting convection, energy losses and particle losses due to nuclear interactions, the propagation of CR protons in the Galaxy is given by the equation
\begin{equation}
\nabla\cdot(D\nabla N)+Q=\frac{\partial N}{\partial t}
\end{equation} 
where $N(\textbf{r},E,t)$ is the differential number density, $E$ is the proton kinetic energy, $D(E)\propto E^a$ with $a=$ constant (positive) is the diffusion coefficient which is assumed to be spatially uniform in the Galaxy and $Q(\textbf{r},E,t)$ is the proton production rate. 

The CR propagation region is assumed to be a cylindrical box bounded in both the radial and vertical directions, and our calculation takes into acount the exact location of the sources with respect to the Earth. Inspite of the fact that the actual spatial distribution of observed SNRs extent as far as $\sim 800 pc$ from the Galactic plane (Stupar et al. 2007), most of the CR propagation studies assume the sources to be uniformly distributed in a thin disc of half-thickness $\sim (150-200) pc$. Such an approximation is valid in the study of global properties of Galactic CRs since majority of the sources are confined within $\sim 200 pc$ from the plane. But, in studies like the present one where the effects of nearby discrete sources are discussed, the actual position of the sources should be considered since, for example, for the same source distance $r_i(x_i,y_i,z_i)$ we expect to see different CR fluxes at different source heights due to the presence of the vertical halo boundary. Our calculation will also assume that the Sun is located on the Galactic plane since our Solar system is only $\sim 15 pc$ away from the plane (Cohen 1995).      

The Green's function $G(\textbf{r},\textbf{r}^\prime,t,t^\prime)$ of Eq. (1), i.e. the solution for a $\delta$-function source term $Q(\textbf{r},t)=\delta(\textbf{r}-\textbf{r}^\prime)\delta(t-t^\prime)$ can be found so that the general solution can be obtained as
\begin{equation}
N(\textbf{r},E,t)=\int^{\infty}_{-\infty}d\textbf{r}^\prime\int^t_{-\infty}dt^\prime G(\textbf{r},\textbf{r}^\prime,t,t^\prime)Q(\textbf{r}^\prime,E^\prime,t^\prime)
\end{equation}
Since the CR particles are assumed to be liberated at time $t=t^\prime$, the equation for $G(\textbf{r},\textbf{r}^\prime,t,t^\prime)$ at $t>t^\prime$ becomes simply
\begin{equation}
\nabla\cdot(D\nabla G)=\frac{\partial G}{\partial t}
\end{equation}
Eq. (3) is solved using the proper boundary conditions and the continuity equations. While solving, we consider the origin to be located at $(x_i,y_i,z_i)$ from the Galactic center. Note that later on this point will represent the actual position of the source with respect to the observer. Then, the CR density at a point $(x>0,y>0,z>0)$ due to a point source [which is positioned at $(x_i,y_i,z_i)$ from the Galactic center] with age $t$, is obtained using Eq. (2) as
\begin{eqnarray}
N(x,y,z,E,t)=\frac{q(E)}{R^2H}\displaystyle\sum_{j=1}^{\infty}\left\lbrace sin\left(\frac{j\pi(R-x_i)}{2R}\right)sin\left(\frac{j\pi(R-x_i-|x|)}{2R}\right)exp\left[-\frac{j^2\pi^2D(t-t_0)}{4R^2}\right]\right\rbrace\nonumber\\
\times \displaystyle\sum_{k=1}^{\infty}\left\lbrace sin\left(\frac{k\pi(R-y_i)}{2R}\right)sin\left(\frac{k\pi(R-y_i-|y|)}{2R}\right)exp\left[-\frac{k^2\pi^2D(t-t_0)}{4R^2}\right]\right\rbrace\nonumber\\
\times \displaystyle\sum_{n=1}^{\infty}\left\lbrace sin\left(\frac{n\pi(H-z_i)}{2H}\right)sin\left(\frac{n\pi(H-z_i-|z|)}{2H}\right)exp\left[-\frac{n^2\pi^2D(t-t_0)}{4H^2}\right]\right\rbrace
\end{eqnarray}
where $R$ and $H$ represent the radial and the vertical boundaries of the Galaxy respectively. The solution at $(x<0,y<0,z<0)$ is obtained by just replacing $(x_i,y_i,z_i)$ with $(-x_i,-y_i,-z_i)$ in Eq. (4). The proton flux can be calculated using $I(E)\approx (c/4\pi)N(E)$, where $c$ is the velocity of light and the source spectrum $q(E)$ is taken as
\begin{equation}
q(E)=k(E^2+2Em_p)^{-(\Gamma+1)/2}(E+m_p)
\end{equation}
in which $m_p$ is the proton mass energy and $k$ is the normalization constant. The source spectral index $\Gamma$ is chosen such that $\Gamma + a=2.73$, the observed proton spectral index (Haino et al. 2004).

For very large radial boundary $(R=\infty)$, the solution of Eq. (1) at $z>0$ can be written as
\begin{equation}
N(x,y,z,E,t)=\frac{q(E)}{4\pi D(t-t_0)H}exp\left[-\frac{(x^2+y^2)}{4D(t-t_0)}\right]
\displaystyle\sum_{n=1}^{\infty}\left\lbrace sin\left(\frac{n\pi(H-z_i)}{2H}\right)sin\left(\frac{n\pi(H-z_i-|z|)}{2H}\right)exp\left[-\frac{n^2\pi^2D(t-t_0)}{4H^2}\right]\right\rbrace
\end{equation} 
Fig. 1 compares the proton flux at the Galactic Center given by Eq. (4) with that of Eq. (6) for an SNR-like source located at $(1,0,0)kpc$ away from the Center with an age $t=2\times 10^4yr$. The results of Eq. (4) at $R=(2,3,10)kpc$ are shown by the thin solid, dashed and dotted lines respectively. The thick solid lines represent the unbounded solution given by Eq. (6) (i.e. the solution for $R=\infty$). The calculations are done at $H=1kpc$ and at $H=3kpc$ assuming $t_0=0$, represented by the left- and right-hand figures respectively. The diffusion coefficient is taken as $D(E)=2\times 10^{28}(E/5GeV)^{0.6}cm^2s^{-1}$ for $E>5 GeV$, where $E$ is in GeV (Engelmann et al. 1990) and the injected protons are assumed to carry $10$ percent of the total explosion energy of $\sim 10^{51}ergs$. The figures clearly show that, for sources near to the observer, the solution of Eq. (4) can be very well approximated by the much simpler unbounded solution for any value of $H$ if $R> 3kpc$. For example, the results at $R=10 kpc$ exactly coincide with the $R=\infty$ lines. Therefore, considering the fact that our solar system is positioned at a distance of $\sim 8.5 kpc$ from the Galactic center and that the Galactic radius extends as far as $\sim 20 kpc$, the effect of the radial boundary $R$ on the observed CRs should be negligible at least for those sources that can give appreciable density fluctuations at the Earth, i.e. for those sources located within $\sim 1.5kpc$ from the Earth (see Thoudam 2006a). In the following sections where we study the effect of nearby SNRs on the observed CRs, we will therefore adopt the simpler Eq. (6) instead of the complicated Eq. (4).    
\begin{figure}
\centering
\includegraphics*[width=0.28\textwidth,angle=270,clip]{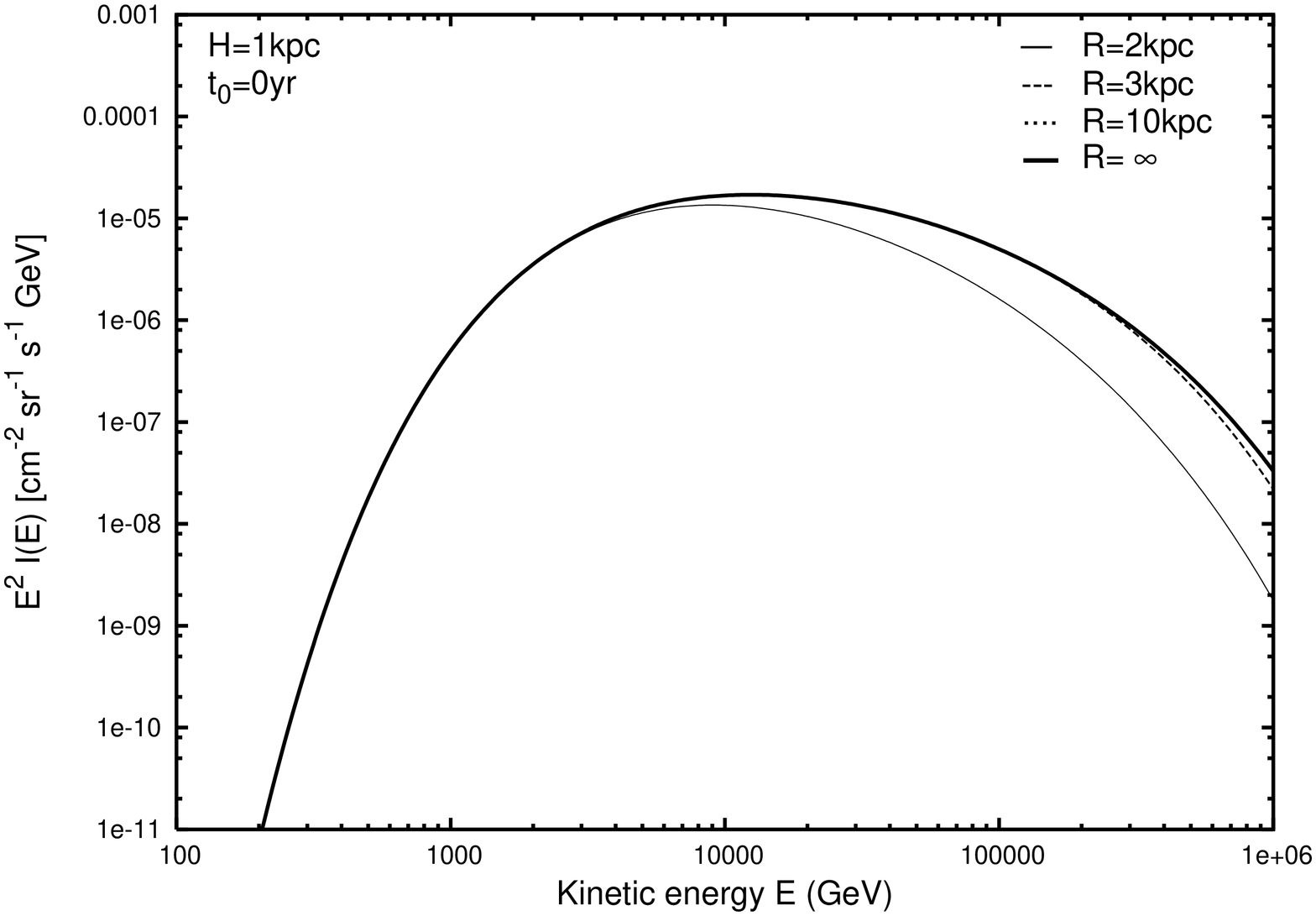}
\includegraphics*[width=0.28\textwidth,angle=270,clip]{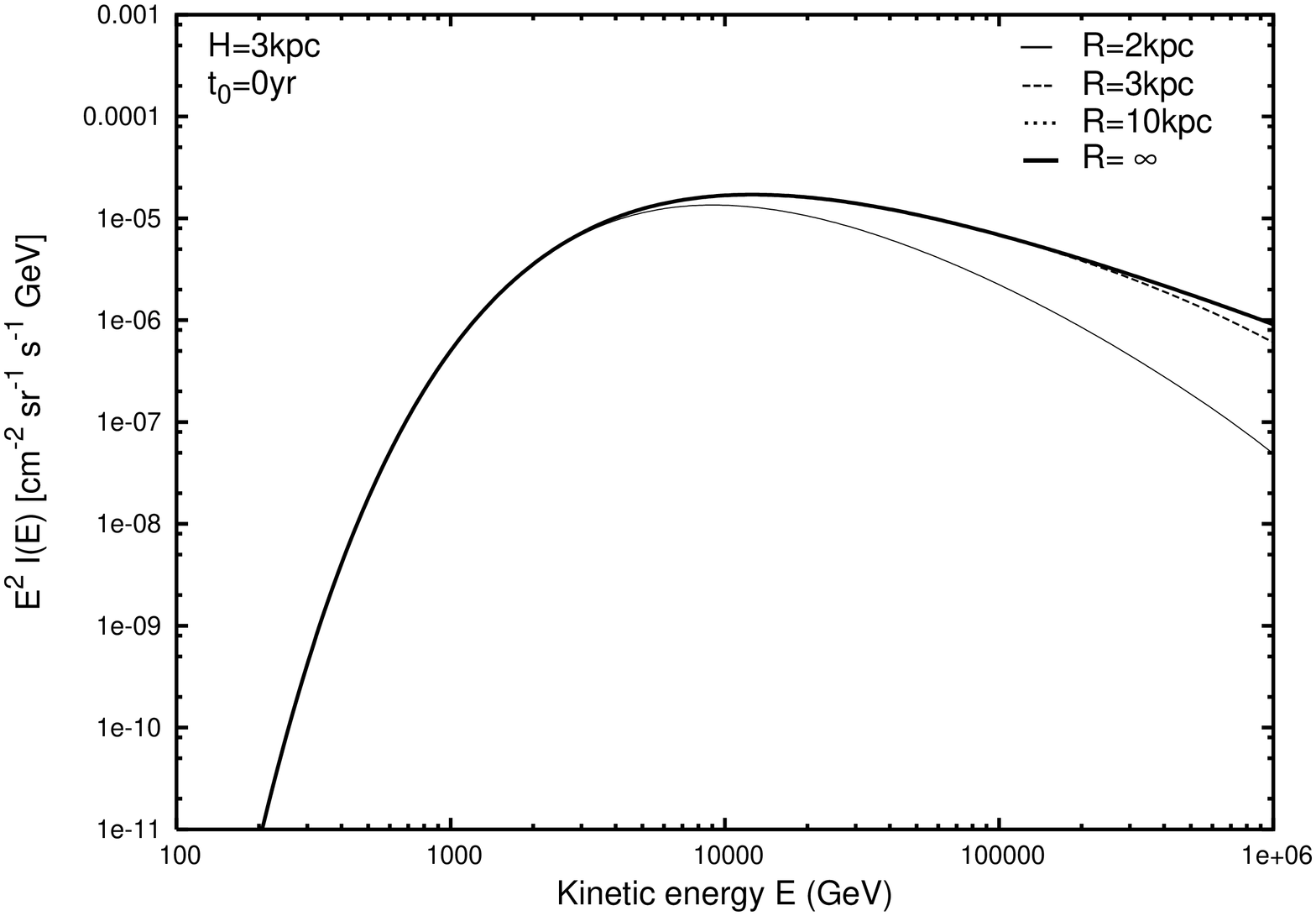}
\caption{\label {fig1} CR proton flux at the Galactic center due to an SNR-like source with age $t=2\times 10^4 yr$ located at $(1,0,0)kpc$ away from the center. The thin solid line, dashed and dotted lines are the fluxes calculated using Eq. (4) for $R=(2,3,10)kpc$ respectively. The thick solid line represents the flux calculated for the boundaryless case ($R=\infty$) using Eq. (6). The calculation assumes $t_0=0$ and $D(E)\propto E^{0.6}$. \textit{Left} : For $H=1kpc$. \textit{Right} : $H=3kpc$. From the figures, it can be seen that for any value of $H$ Eq. (4) can be well approximated by the much simpler boundaryless solution if $R> 3kpc$.}
\end{figure}
\section {CR anisotropy}
Knowing the CR density at a point $(x,y,z)$ away from a source of age $t$, the single source anisotropy amplitude in the diffusion approximation can be calculated using (Mao $\&$ Shen 1972)
\begin{equation}
\delta_i=\frac{3D}{c}\frac{|\nabla N_i|}{N_i}
\end{equation}
where $N_i$ is given by Eq. (6) for a point source $i$ located at $(x_i,y_i,z_i)$ from the Earth. The total anisotropy parameter at the Earth due to a number of nearby discrete sources in the presence of an isotropic CR background is given by (Paper I)
\begin{equation}
\delta=\frac{\displaystyle\sum_{i} I_i\delta_i \hat{r}_i.\hat{n}_m}{I_T}
\end{equation} 
where the summation is over the nearby discrete sources. $\hat{r}_i$ denotes the direction of the source $i$ giving a flux $I_i$ and $\hat{n}_m$ denotes the direction of maximum intensity. $I_T=1.37(E/GeV)^{-2.73} cm^{-2}s^{-1}sr^{-1}GeV^{-1}$ represents the total observed flux of CR protons above $\sim 10GeV$ (Haino et al. 2004). The phase of the anisotropy is taken as the direction of maximum intensity. Therefore, the anisotropy $\delta$ as well as the phase at an energy $E$ depends on the age and distance of the nearby sources, and may be determined by different sources at different energy intervals. However, in the case of a single source dominance, the total anisotropy $\delta$ is given by $\delta=(I_m/I_T)\delta_m$, where $m$ denotes the source giving the maximum flux at the Earth. 
\section{Comparison with the results of Paper I}
\begin{figure}
\centering
\includegraphics*[width=0.28\textwidth,angle=270,clip]{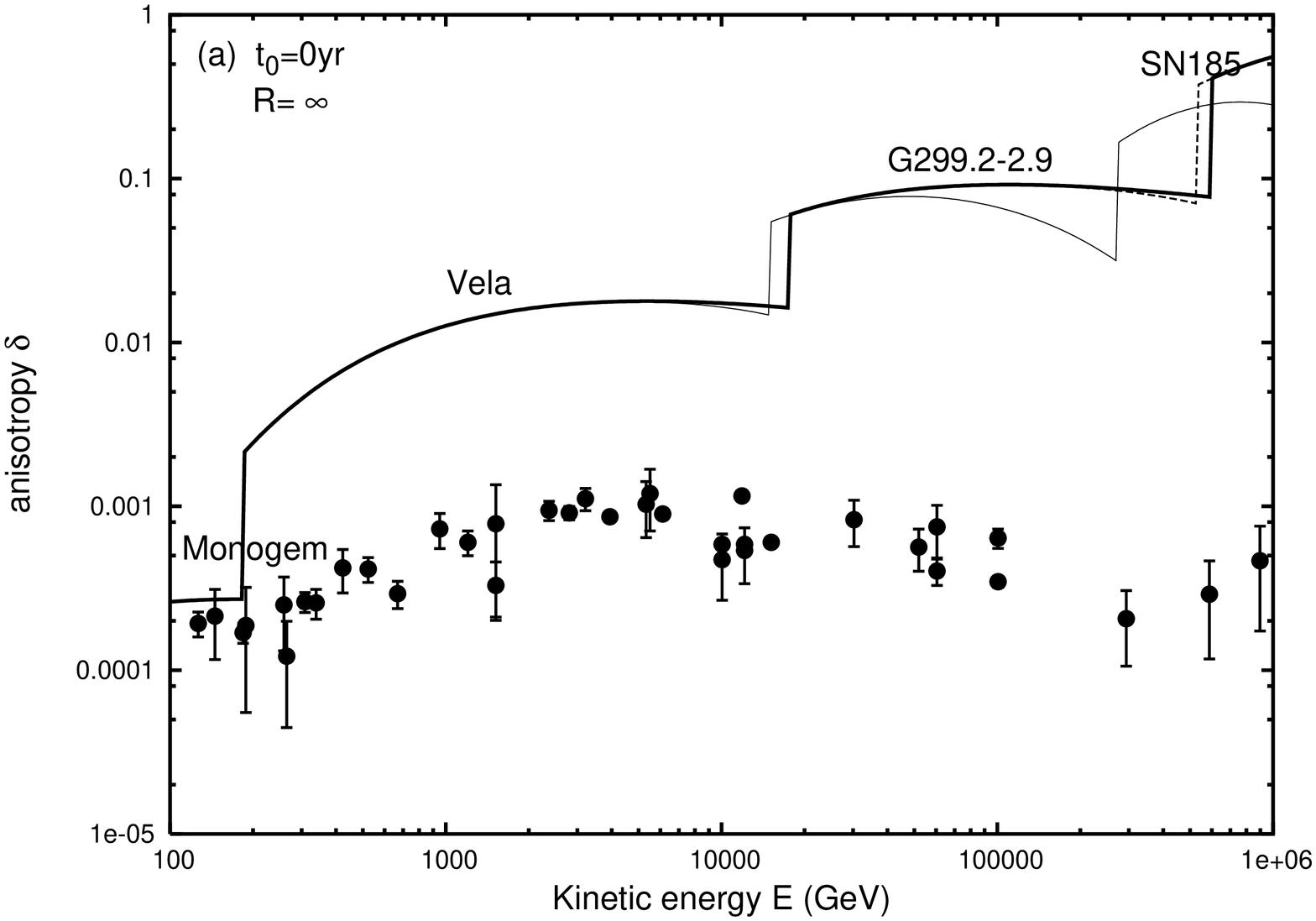}
\includegraphics*[width=0.28\textwidth,angle=270,clip]{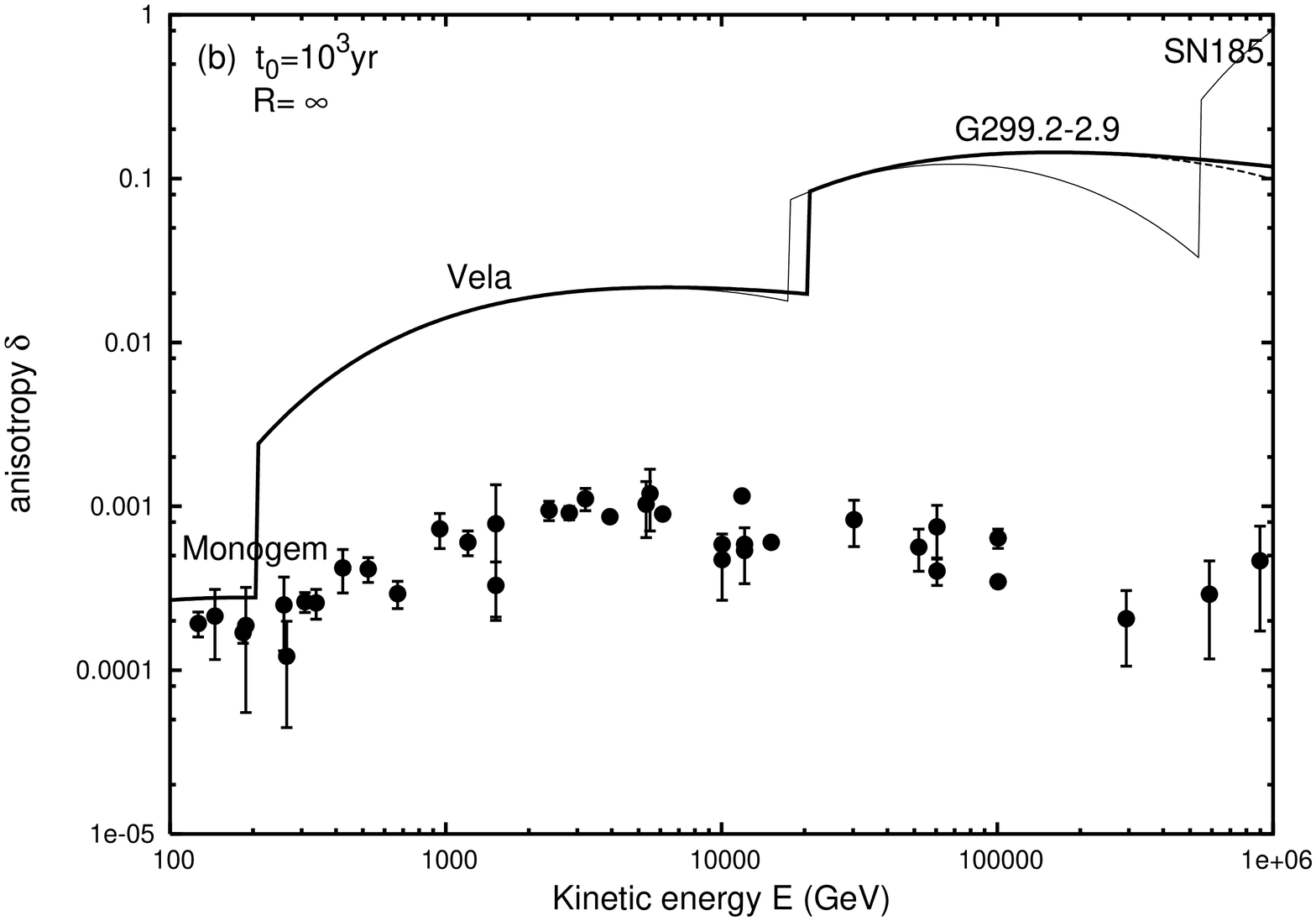}
\includegraphics*[width=0.28\textwidth,angle=270,clip]{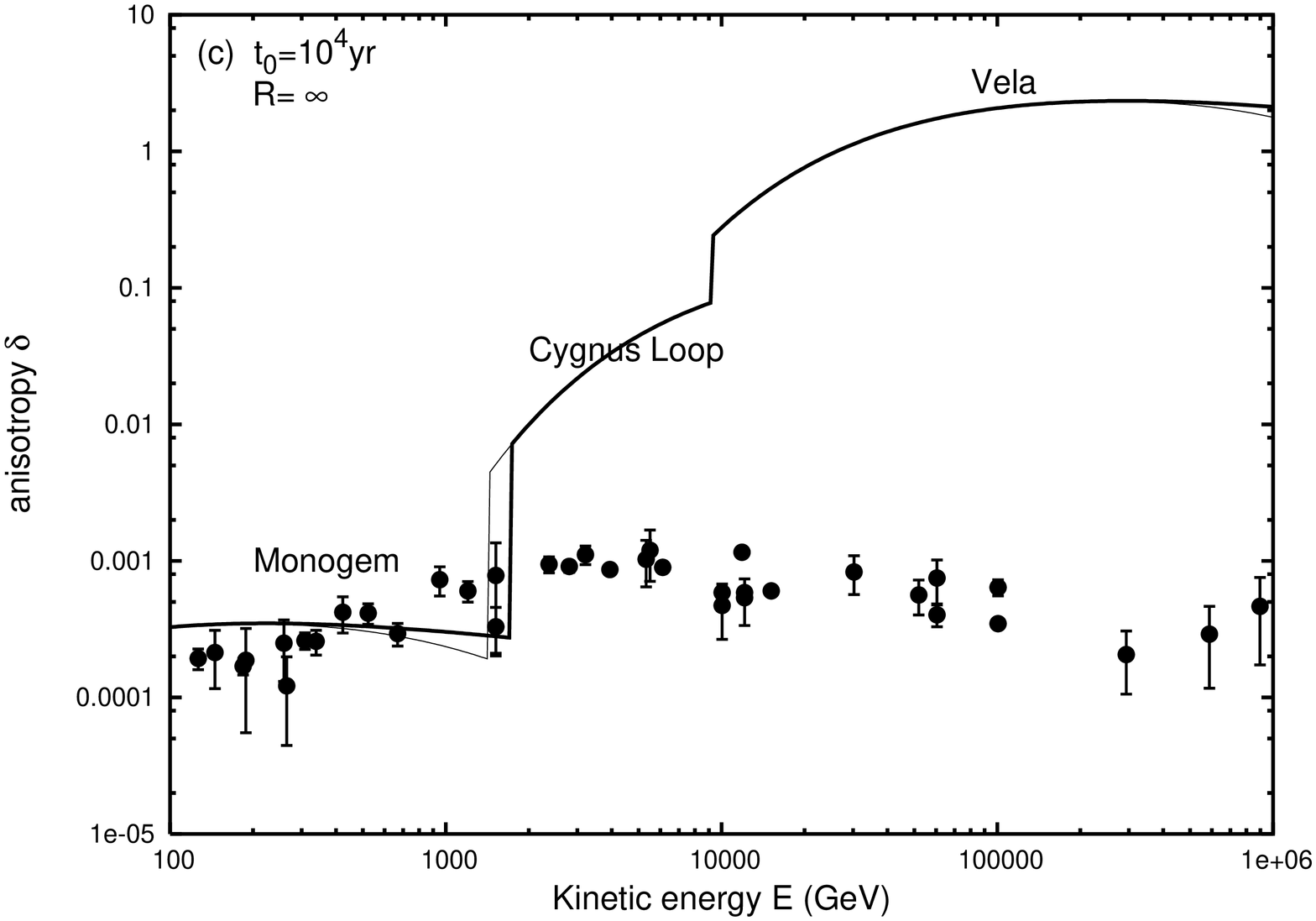}
\includegraphics*[width=0.28\textwidth,angle=270,clip]{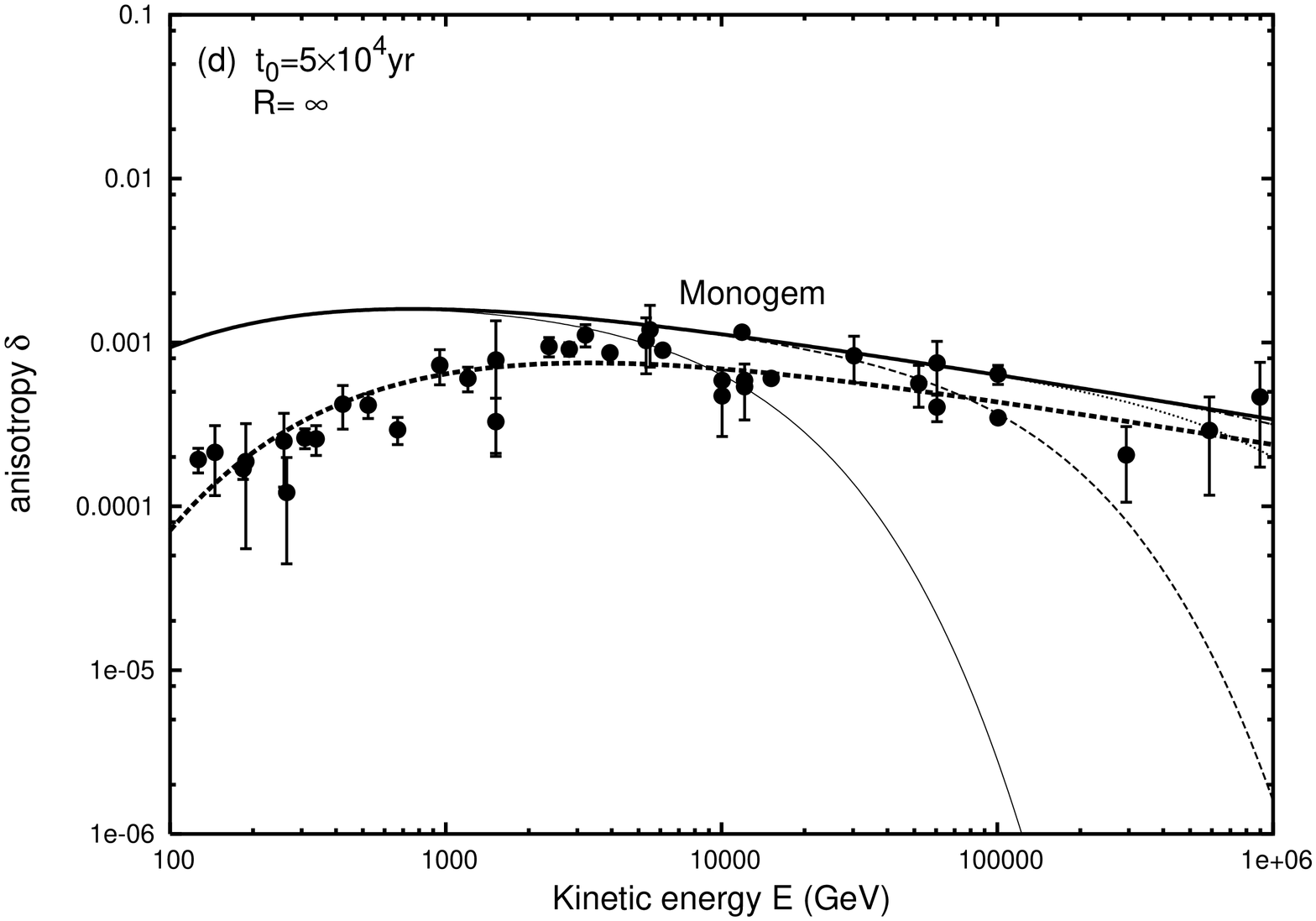}
\caption{\label {fig1} CR anisotropy at the Earth due to nearby known SNRs assuming burst-like injection model. Figs. (a), (b), (c) $\&$ (d) are the results obtained for $t_0=(0,10^3,10^4$ $\&$ $5\times 10^4) yr$ respectively. Thin solid lines represent the results of Eq. (8) for $H=0.5kpc$, the thin dashed lines for $H=1kpc$, dotted lines for $H=2kpc$ and the dot-dashed lines for $H=3kpc$. The thick solid lines are the results of Paper I i.e. for $H=\infty$. Figs. (a), (b) $\&$ (c) show that different sources determine the anisotropy at different energy intervals. These are marked by the source names along the lines. In Fig. (d), the Monogem SNR solely determine the anisotropy in the whole energy range. The thick dashed line is the best fit result, in the case of a single source dominance, calculated assuming infinite boundaries. Data points are taken from the compilation of various results given in Erlykin $\&$ Wolfendale 2006.}
\end{figure}
In this section, we will try to investigate whether the presence of a halo boundary can affect the anisotropy at the Earth due to nearby sources. For that, we consider the $13$ known SNRs located within $1.5kpc$ from the Earth as listed in Table 1 of Paper I. The total anisotropy due to these SNRs is calculated using Eq. (8) for different $H$ values at different $t_0$'s.
 
Fig. 2 shows the comparison of the anisotropies calculated in the present work with those obtained in Paper I for the burst-like particle injection model. The data points are taken from the compilation of various experiments given in Erlykin $\&$ Wolfendale (EW) 2006. Figs 2(a), (b), (c) and (d) are the results obtained for $t_0=(0,10^3,10^4$ $\&$ $5\times 10^4) yr$, respectively. The thin solid lines represent the results of Eq. (8) for $H=0.5kpc$, the dashed lines are for $H=1kpc$, the dotted lines are for $H=2kpc$ and the dot-dashed lines are for $H=3kpc$. The thick solid lines are the results of Paper I which were obtained assuming $H=\infty$ [Eq. 11 of Paper I]. In Figs 2(a)$-$(c), different sources determine the anisotropy at different energy ranges. These are marked by the source names along the lines. It can be seen that the results for $H=0.5kpc$ show a noticeable deviation from the $H=\infty$ lines, while those for $H=1kpc$ show a very slight deviation. The results for other higher $H$- values almost overlap with the $H=\infty$ lines and are not easily visible in the figures. This shows that, for the particle release time of $t_0 \lesssim 10^4 yr$, the halo height effect on the local SNR contribution to the observed CR anisotropy is almost negligible if $H>1 kpc$. However, the situation is somewhat different in Fig. 2(d) where the calculations are performed at $t_0=5\times 10^4yr$. Note that this value of particle injection time is that at which the model calculated anisotropy values are close to the observed data (see the results of Paper I). The anisotropy here is determined solely by the Monogem SNR in the whole energy range considered here, and only those results for $H\lesssim 2 kpc$ show considerable  variation from the $H=\infty$ line. The results for $H>2 kpc$ show a negligible deviation. Combining all the results of Fig. 2, we can finally conclude that the effect of the halo boundary of our Galaxy on the local SNR contribution to the observed CR anisotropy is negligible as long as the boundary is greater than $\sim 2 kpc$. In the next section, we will combine this result along with the halo heights obtained by several authors to discuss the importance of $H$ in the anisotropy study due to local sources.         

\section{Discussions and conclusions}
The effect of the nearby CR sources is considered as one of the important effects that can give rise to the observed CR anisotropy at the Earth. However, the calculation of CR fluxes from any type of source in the Galaxy essentially requires the use of the proper geometry of the Galaxy as well as the actual position of the source with respect to the observer. Since our Galaxy has a cylindrical geometry with the radius much larger than the height, the radial boundary is found to have a negligible effect on the CR density and hence the geometry can be approximated by an infinite radius with a finite vertical height. Furthermore, this study has found that the effect of the vertical halo boundary on the local SNR contribution to the CR anisotropy is negligible if $H>2kpc$. 

Fig. 2 shows the effect of the halo height on the CR anisotropy due to nearby known sources for different particle injection times. Among the 13 SNRs considered, only Monogem, Vela, G299.2-2.9, SN185 and Cygnus Loop are found to determine the anisotropy at different energy intervals. Also, all of them except SN185 (with  $r=0.95kpc$) have distances $r\lesssim 0.5kpc$. This shows that only the nearest sources mainly determine the anisotropy as expected, and hence this results in a negligible halo height effect for $H>2 kpc$. It is also worth mentioning that the vertical heights of the dominant sources above the Galactic plane are found to be less than $\sim 60 pc$ which is much less than the halo heights $(H\geqslant 0.5kpc)$ considered here. 

The actual value of the halo height of our Galaxy is not exactly known. Its value is generally obtained along with other propagation parameters using the observed CR data like the secondary$/$primary ratios, CR density distribution etc. But, the values obtained from the same experimental data are different for different CR propagation models. Webber, Lee $\&$ Gupta (1992) had obtained a value of $H<4 kpc$ using diffusion-convection model. Lukasiak et al. (1994) had obtained $H=2.8^{+1.2}_{-0.9} kpc$ using the Webber et al. (1992) model without convection. Webber $\&$ Soutoul (1998) obtained $H=(2-4)kpc$ and $H=(2-3)kpc$ using the diffusion and  Monte Carlo models respectively. Other results like those of Freedman et al. (1980) and Ptuskin $\&$ Soutoul (1998) obtained $H\geqslant 7.8 kpc$ and $H=4.9^{+4}_{-2}kpc$ respectively. A completely numerical approach using more  realistic physical conditions of the Galaxy determined a value of $H>4kpc$ for the diffusion-convection model and $H=(4-12)kpc$ for the re-acceleration model (Strong $\&$ Moskalenko 1998). These results are found to be consistent with the observations of Galactic radio emission structure at 408 MHz which indicate the presence of a thick radio disk with full equivalent width of $(2.3\pm 0.2)kpc$, $(3.6\pm 0.4)kpc$ and $(6.3\pm 0.7)$ in the Galactic radial range of $(0-8)kpc$, $(8-12)kpc$ and $(12-20)kpc$ respectively (Beuermann et al. 1985), but such a wide range of values makes the Galactic halo height a very uncertain parameter in CR propagation studies. However, since most of the values obtained are found to have $H\gtrsim 2kpc$, the conclusion given in the previous section suggests that the study of local CRs due to nearby SNRs can be carried out without having much information on $H$. This is because the effect of the nearest sources $(r\lesssim 0.5kpc)$ dominates over the influence of the other nearby sources and the CR fluxes from these sources are almost independent of the halo boundary for $H>2kpc$ as discussed before. Hence, the study of the effect of local sources on the CR anisotropy at the Earth can be done using the much simpler three-dimensional unbounded solution. 

For the infinite boundary case, if a single source dominates the anisotropy in the whole energy range as in Fig. 2(d), the total anisotropy follows an energy dependence of the form $\delta\propto E^{-a/2}$ in the high energy regime (Paper I), which for $a=0.6$ goes as $\delta\propto E^{-0.3}$. Such a decrease with energy is in fact observed in the high energy anisotropy data somewhere above $E\sim 4\times 10^3GeV$ upto around $3\times 10^5 GeV$. Moreover, the increase in anisotropy from $\sim (10^{-4}-10^{-3})$ in the $(10^2-3\times 10^3)GeV$ energy range can also be possibly explained by a proper choice of $(r,t,t_0)$ or rather $(r,\Delta_t=t-t_0)$ for the single dominant source. We try to estimate the physical parameters of such a source that best fit the data. The best-fitting parameters are found to be $r=(0.570\pm 0.023)kpc$ and $\Delta_t=(5.343\pm 0.224)\times 10^4 yr$, and the best-fitting line is shown as the thick dashed line in Fig. 2(d). Thus, for $t_0=0$ the source should have an age of $t=\Delta_t$. However, it is possible to obtain a number of $(t,t_0)$ combinations which equally fit the data, all of them giving the same value of $r$ and $\Delta_t$. Therefore, the present study only gives an estimate of the distance to the single dominant source; it does not give any precise information on the age and the particle release time of the source. It should be noted that it is not the individual $(t,t_0)$ values that determine the contribution of the source, but the propagation time $\Delta_t$ of the particles after their release from the source. We can determine the best-fitting $t$- value only if we know $t_0$, but the value of $t_0$ is not exactly known. It may even be that $t_0$ is an energy-dependent parameter, i.e. particles with different energies emitted at different times. Studies based on diffusive shock acceleration in SNRs have shown that the highest energy particles start leaving the source region already at the beginning of the Sedov phase (Berezhko et al. 1996), but the major fraction of accelerated CRs remain confined for almost around $10^5yr$ for an interstellar medium (ISM) hydrogen atom density of $n_H=1 cm^{-3}$. This implies that for the local ISM which has $n_H\sim 1 cm^{-3}$ (see e.g. Thoudam 2006b and references therein), if a single source determines the whole anisotropy, the source should have a characteristic age of $\sim 1.5\times 10^5 yr$. Unfortunately, there is no nearby known SNR with such an age located at $r\sim 0.57 kpc$. However, it is quite possible that the single dominant source may be an \textit{undetected} old SNR. In fact, studies assuming adiabatic phase in SNR evolution have shown that the surface brightness of an SNR of age $\sim 10^5$ yrs lies below the detection limit of radio telescopes (Leahy $\&$ Xinji 1989). The present result is further supported by the  fact that almost all the nearby sources are quite young with estimated ages less than $10^5 yr$ (the generally accepted particle release time), and they might not have released the CRs into the local ISM. In addition, the possiblity that some of the observed features of CRs may be due to \textit{undetected} nearby sources cannot be simply ignored. 

The single-source explanation of the observed CR properties can also be found in some earlier works (e.g. EW 2000 and references therein; EW 2006, etc.), but in a somewhat different context. EW 2000 claimed that the knee in the CR spectrum at $E\sim 3 PeV$ can be attributed to the presence of a single recent supernova (as yet unidentified) in the local region. On the other hand, EW 2006 tried to explain the rise in the anisotropy amplitude as well as the change in its phase near the knee using a single source exploded in the direction from the Sun downward of the main CR flux, which are predominantly coming from the inner Galaxy. The latter study considered the source parameters as similar to those of the Monogem SNR. Although the single source idea has not been readily accepted by the CR community, at the same time there is no reason why it should be just neglected. The present study even points out one more observed property of CRs that can possibly be explained by the single source model.


\begin{thebibliography}{99}
\bibitem{}
Berezhko, E. G., Yelshin V.K. $\&$ Ksenofontov L.T. 1996, J. Exp. Theor. Phys., 82, 1
\bibitem{}
Beuermann, K., Kanbach, G., $\&$ Berkhuijsen, E. M. 1985, A$\&$A, 153, 17
\bibitem{}
Cohen, M., 1995, ApJ, 444, 874
\bibitem{}
Engelmann, J. J., Ferrando, P., Soutoul, A., Goret, P., $\&$ Juliusson, E. 1990, A$\&$A, 233, 96
\bibitem{}
Erlykin, A. D., $\&$ Wolfendale, A. W. 2000, A$\&$A, 356, L63 
\bibitem{}
Erlykin, A. D., $\&$ Wolfendale, A. W. 2006, Astropart. Phys., 25, 183
\bibitem{}
Freedman, I., Kearsey, S., Osborne, J. L., $\&$ Giler, M. 1980, A$\&$A, 82, 110
\bibitem{}
Guillian, G., et al. 2007, Phys. Rev. D, 75, 062003
\bibitem{}
Haino, S., et al. 2004, Phys. Lett. B594, 35
\bibitem{}
Leahy, D. A., $\&$ Xinji, W. 1989, PASP, 101, 607
\bibitem{}
Lukasiak, A., Ferrando, P., McDonald F. B., $\&$ Webber, W. R. 1994, ApJ, 423, 426
\bibitem{}
Mao, C. Y., $\&$ Shen, C. S. 1972, Chinese J. Phys., 10, 16
\bibitem{}
Ptuskin, V. S., $\&$ Soutoul, A. 1998, A$\&$A, 337, 859
\bibitem{}
Strong, A. W., $\&$ Moskalenko, I. V. 1998, ApJ, 509 212
\bibitem{}
Stupar, M., Filipovi$\acute{c}$, M. D., Parker, Q. A., White, G. L., Pannuti, T. G., $\&$ Jones, P. A. 2007, Ap$\&$SS, 307, 423
\bibitem{}
Thoudam, S. 2006a, MNRAS, 370, 263
\bibitem{}
Thoudam, S. 2006b, Astropart. Phys., 25, 328
\bibitem{}
Thoudam, S. 2007, MNRAS, 378, 48 (Paper I)
\bibitem{}
Webber, W. R., Lee, M. A., $\&$ Gupta, M. 1992, ApJ, 390, 96
\bibitem{}
Webber, W. R., $\&$ Soutoul, A. 1998, ApJ, 506, 335
\end{thebibliography}
\end{document}